\documentclass[aps,prd,twocolumn,a4paper,superscriptaddress,floatfix]{revtex4-1}
\usepackage{graphicx}
\usepackage{hyperref}
\usepackage{amssymb}
\usepackage{natbib}
\usepackage{color}
\usepackage[dvipsnames]{xcolor}
\usepackage{amsmath}
\begin{document}
\newcommand{\be}{\begin{equation}}
\newcommand{\ee}{\end{equation}}
\newcommand{\bq}{\begin{eqnarray}}
\newcommand{\eq}{\end{eqnarray}}
\newcommand{\bw}{\begin{widetext}}
\newcommand{\ew}{\end{widetext}}
\newcommand{\bsq}{\begin{subequations}}
\newcommand{\esq}{\end{subequations}}
\newcommand{\bc}{\begin{center}}
\newcommand{\ec}{\end{center}}

\newcommand\lsim{\mathrel{\rlap{\lower4pt\hbox{\hskip1pt$\sim$}}
 	\raise1pt\hbox{$<$}}}
\newcommand\gsim{\mathrel{\rlap{\lower4pt\hbox{\hskip1pt$\sim$}}
 	\raise1pt\hbox{$>$}}}

\def\app#1#2{%
  \mathrel{%
    \setbox0=\hbox{$#1\sim$}%
    \setbox2=\hbox{%
      \rlap{\hbox{$#1\propto$}}%
      \lower1.1\ht0\box0%
    }%
    \raise0.25\ht2\box2%
  }%
}
\def\approxprop{\mathpalette\app\relax}

%%%%%%%%%%%%%%%%%%%%%%%%%%%%
\title{Deviations from the von Laue condition: Implications for the on-shell Lagrangian of particles and fluids}
\author{S. R. Pinto}
\email{up202004386@edu.fc.up.pt}
\affiliation{Departamento de Física e Astronomia, Faculdade de Ci\^encias, Universidade do Porto, Rua do Campo Alegre, 4169-007 Porto, Portugal}
\affiliation{Instituto de Astrof\'{\i}sica e Ci\^encias do Espa\c co, CAUP, Rua das Estrelas, 4150-762 Porto, Portugal}
\author{P. P. Avelino}
\email{pedro.avelino@astro.up.pt}
\affiliation{Departamento de Física e Astronomia, Faculdade de Ci\^encias, Universidade do Porto, Rua do Campo Alegre, 4169-007 Porto, Portugal}
\affiliation{Instituto de Astrof\'{\i}sica e Ci\^encias do Espa\c co, CAUP, Rua das Estrelas, 4150-762 Porto, Portugal}

\begin{abstract}

According to the von Laue condition, the volume integral of the proper pressure inside isolated particles with a fixed structure and finite mass vanishes in the Minkowski limit of general relativity. In this work, we consider a simple illustrative example: nonstandard static global monopoles with finite energy, for which the von Laue condition is satisfied when the proper pressure is integrated over the whole space. We demonstrate, however, that the absolute value of this integral, when calculated up to a finite distance from the center of the global monopole, generally deviates from zero by no more than the energy located outside the specified volume (under the assumption of the dominant energy condition). Furthermore, we find that the maximum deviation from unity of the ratio between the volume averages of the on-shell Lagrangian and the trace of the energy-momentum tensor cannot exceed three times the outer energy fraction. Extending these results to real particles, we demonstrate that these constraints generally hold for finite-mass systems with fixed structure, including stable atomic nuclei, provided the dominant energy condition is satisfied. Specifically, we show that, except in extremely dense environments with energy densities comparable to that of the particles themselves, the volume average of the aforementioned ratio must be extremely close to unity. Finally, we discuss the broader implications of our findings for the form of the on-shell Lagrangian of real fluids, which is often a crucial element for accurately modeling the dynamics of the gravity and matter, especially in scenarios involving nonminimal couplings to other matter fields or gravity. We find that, in general, the ideal gas on-shell Lagrangian provides an accurate approximation of the true on-shell Lagrangian, even for nonideal gases with significant interparticle interactions.

\end{abstract}
\date{\today}
\maketitle

%%%%%%%%%%%%%%%%%%%%%%%%%%%%%
\section{Introduction \label{sec1}}

Perfect fluids provide a continuous, high-level description of a medium in the absence of viscosity and heat conduction. In many cases, they serve as an accurate model for the material content of the Universe, as their properties naturally emerge from the collective behavior of vast ensembles of particles, including baryons, photons, neutrinos, and dark matter. Their proper density, proper pressure, and four-velocity are defined over volume elements containing numerous particles --- microscopic fluctuations are smoothed out, but the macroscopic dynamics is preserved. This makes perfect fluids a fundamental tool in astrophysics and cosmology.

For minimally coupled fluids this volume-averaged high-level description makes no explicit reference to the fluid on-shell Lagrangian.  Nevertheless, perfect fluids are often described in terms of a Lagrangian formulation \cite{Schutz:1970my, Ray:1972, Schutz:1977df, Taub:1978, Matarrese:1984zw, Brown:1992kc,  Andersson:2006nr,  Minazzoli:2012md, Ferreira:2020fma}, which provides a useful framework for studying their macroscopic behavior while neglecting the intricate details of the microscopic dynamics taking place at the particle level. Some of these studies have shown that equivalent Lagrangian formulations can be constructed, where the perfect fluid on-shell Lagrangian takes the form  $\mathcal{L}_{\text{on-shell}}=p$ or $\mathcal{L}_{\text{on-shell}}=-\rho$ [or, more generally, $\mathcal{L}_{\text{on-shell}}=-C\rho + (1-C)p$ \cite{Brown:1992kc,LopesdeAzevedo:2022fex}, where $C$ is a real constant]. This is often  misinterpreted within the community as an arbitrary freedom to choose the on-shell Lagrangian of a perfect fluid, rather than a consequence of the simplifications inherent in these Lagrangian formulations (which neglect microscopic effects).

This issue becomes particularly significant in the presence of nonminimal couplings \cite{Wetterich:1994bg, Amendola:1999er, Zimdahl:2001ar, Farrar:2003uw, Capozziello_2011, OLMO_2011, Clifton_2012, Planck:2014ylh, Berti_2015, Planck:2015bue, Nojiri_2017} to other fluids or fields (including the gravitational field), where the dynamics of both the perfect fluid and of the other coupled fluids or fields often depend explicitly on the on-shell Lagrangian of the perfect fluid \cite{Harko_2010, Harko_2011, Kat_rc__2014, Haghani_2013, Ludwig_2015, Harko_2018, Bahamonde_2018}. Although no universal on-shell Lagrangian exists for a fluid (even for a perfect fluid), it has been demonstrated that $\mathcal{L}_{\text{on-shell}}=T$ for an ideal gas ($T$ is the trace of the energy-momentum tensor) \cite{Avelino:2018qgt, Avelino:2018rsb, Avelino_2022}, thus implying that the two most common assumptions in the literature regarding the on-shell Lagrangian of a perfect fluid ($\mathcal{L}_{\text{on-shell}}=p$ and $\mathcal{L}_{\text{on-shell}}=-\rho$) are not valid for an ideal gas (except in the special case of dust, where $\mathcal{L}_{\text{on-shell}}=-\rho$). This result heavily relies on the von Laue condition \cite{Laue:1911} (see also \cite{Avelino:2018qgt,Giulini:2018tuw,Avelino:2023rac,Avelino:2024pcl}), which states that the volume-averaged pressure inside stable, static, compact objects with a negligible self-induced gravitational field vanishes. However, the von Laue condition does not account for deviations that may arise in dense environments due to interparticle interactions. In this work, we overcome this limitation by examining how interactions between particles and the consequent breakdown of the ideal gas approximation may affect this condition. Specifically, we explore how these interactions influence both the validity of the von Laue condition and the form of the on-shell Lagrangian for particles and their associated fluids. To this end, we begin by considering the illustrative case of nonstandard static global monopoles, or more specifically, global K-monopoles, as a toy model for finite-mass particles (see \cite{Babichev_2006,Avelino_2011} for more details on global K-defects). 

The outline of this paper is as follows. In Sec. \ref{GM}, we introduce a scalar field model featuring a nonstandard kinetic term and an O(3)-symmetric potential that admits global monopole solutions, known as K-monopoles. These solutions serve as a particle toy model for exploring various properties of real particles. Section  \ref{sec3} presents a derivation of the von Laue condition, an essential constraint for the stability of compact objects with a negligible self-induced gravitational field. In Sec. \ref{sec4}, we describe the analytical framework and numerical methods used to obtain static global K-monopole solutions, while Sec. \ref{sec5} discusses the numerical results, focusing on the dependence of key physical variables on the radial distance from the monopole center for different choices of the nonstandard kinetic term. In Sec. \ref{sec6}, we derive an upper limit for deviations from the von Laue condition which may arise due to interparticle interactions in realistic scenarios. Section \ref{sec7} builds on these results to examine the form of the on-shell Lagrangian for real fluids, considering both minimally coupled fluids and those with nonminimal couplings to other fields or gravity. Finally, in Sec. \ref{concl}, we summarize our findings, highlighting their implications for the dynamics of particles and their associated fluids.

Throughout this work, we adopt the metric signature $[-, +, +, +]$ and use natural units where $c = \hbar = 1$. The Einstein summation convention is employed, meaning that when an index variable appears twice in a single term it implies summation over all possible values of the index. Unless explicitly stated otherwise, Greek indices run over $\{0, 1, 2, 3\}$, while Latin indices are restricted to $\{1, 2, 3\}$.

\section{global K-monopole: a particle toy model \label{GM}}

In this section, we consider a scalar field model with a nonstandard kinetic term and an O(3)-symmetric potential that admits global monopole solutions, referred to as K-monopoles. Later, we will use these solutions to illustrate various properties of real particles.

We begin by considering a real scalar field multiplet $\{ \phi^1, \phi^2, \phi^3 \}$, in a $3+1$-dimensional Minkowski spacetime, described by the Lagrangian
\begin{equation}
\mathcal{L} = K(X) - V(\phi^a)\,,
\label{eq1}
\end{equation}
where 
\begin{equation}
X = -\frac{1}{2} \partial^{\mu}\phi^a \partial_{\mu} \phi^a\,.
\label{eq2}
\end{equation} 
The corresponding energy-momentum tensor is given by
\begin{equation}
T_{\mu \nu} \equiv -\frac{2}{\sqrt{-g}}\frac{\delta(\mathcal L \sqrt{-g})}{\delta g^{\mu \nu}}= g_{\mu \nu} \mathcal{L}+K_{,X}\partial_{\mu}\phi^a \partial_{\nu}\phi^a\,,
\label{eq3}
\end{equation}
where $g = {\rm det}(g_{\mu \nu})$, $g_{\mu \nu}$ are the components of the metric tensor, and a comma in the subscript denotes a partial derivative with respect to $X$. 

For a standard kinetic term $K(X)=X$, Derrick and Hobart's no-go theorem \cite{Derrick_1964,Hobart_1963} rules out static, finite-energy solutions in Minkowski spacetime for more than one spatial dimension. To obtain such solutions, nonstandard kinetic terms are required. For the sake of definiteness, we shall assume that $K(X)=X |X|^{\alpha-1}$, with $\alpha \geq 1$, throughout the paper. We will also consider an O(3)-symmetric scalar potential given by
\begin{equation}
V(\phi^a) = \frac{\lambda}{4} (\phi^a\phi^a-\eta^2)^2\,,
\label{eq4}
\end{equation} 
where $\lambda$ is a dimensionless constant, and $\phi^a\phi^a =\eta^2$ defines the vacuum manifold as a two-sphere. For simplicity, unless otherwise specified, we use units where $\eta = 1$ throughout the paper. 

The equation of motion for the scalar multiplet can be written as
\begin{equation}
K_{,X} \partial^\mu \partial_\mu \phi^a + K_{,XX} \partial_\mu X  \partial^\mu \phi^a - \frac{\partial V}{\partial \phi^a} = 0\,.
\label{eq5}
\end{equation}
To find maximally symmetric static solutions of global K-monopoles,  we use the ansatz
\begin{equation}
\phi^a = \frac{x^a}{r} f(r)\,, 
\label{eq6}
\end{equation}
with $r^2 = x^a x^a$, $f(0)=0$, and $f(r\rightarrow\infty)=1$. Notice that Eq. (\ref{eq6}) implies that $ \phi^a \phi^a = f^2(r)$, so that the minima of $V(\phi^a)$ are at $f=1$. The field equation for $f$ is
\begin{equation}
 \left(f'' + \frac{2}{r}f' - \frac{2}{r^2}f \right)  - (\alpha-1) \frac{X'}{|X|} f' - \frac{1}{\alpha|X|^{\alpha-1}}\frac{dV}{df} = 0\,,
 \label{eq7}
\end{equation}
where
\begin{eqnarray}
X = X(r) &=& -\frac12\left(f'^2 +\frac{2}{r^2} f^2\right)\,, \label{eq8}\\
X'= X'(r) &=& - f'f''-2\frac{ff'r^2-rf^2}{r^4}\, \label{eq9}\\
V = V(f) &=& \frac{\lambda}{4} (f^2 - 1)^2\,, \label{eq10}\\
\frac{dV}{df}&=&\lambda f(f^2-1)\,, \label{eq11}
\end{eqnarray}
and a prime denotes a derivative with respect to $r$.

The proper energy density $\rho(r)$, the proper pressure $p(r)$, and the trace of the energy-momentum tensor $T(r)$, are given, respectively, by
\begin{eqnarray}
\rho(r)&=&T_{00}= -\mathcal{L}_{\rm on-shell} \nonumber \\
&=& |X(r)|^\alpha+\frac{1}{4} \lambda (f(r)^2 - 1)^2 \label{eq12}\,,\\
p(r)&=&  \frac{T_{ii}}{3}= 
\mathcal{L}_{\rm on-shell}(r)+ \frac{2\alpha}{3}|X(r)|^\alpha
\label{eq13}\,,\\
T(r)&=&  -\rho + 3 p=4\mathcal{L}_{\rm on-shell} +2\alpha|X(r)|^\alpha  \,.
\label{eq14}
\end{eqnarray}
At large distances from the global K-monopole center ($r\gg 1$), $f$ approaches unity. This implies that the dominant contribution to $X$ arises from the angular variation of the scalar field,
\begin{equation}
   X(r) \sim -\frac{1}{2} \left(\partial^{\theta}\phi^a \partial_{\theta} \phi^a +\partial^{\varphi}\phi^a \partial_{\varphi} \phi^a \right) \sim -1/r^2\,,
   \label{eq15}
\end{equation}
where $\theta$ and $\varphi$ represent angular spherical coordinates. In this regime, the potential energy density becomes negligible ($V \sim 0$, for $f \sim 1$) and
\begin{equation}
    \rho(r) \sim 1/r^{2\alpha}\,.
\label{eq16}
\end{equation}
The total mass (or energy) of the monopole, 
\begin{equation}
m=4\pi \int_0^\infty \rho(r) r^2 dr \,,
\label{eq17}
\end{equation}
diverges if $\alpha \le 3/2$, consistent with Derrick and Hobart's theorem. For $\alpha > 3/2$ the total mass converges. In this case, the mass of the global K-monopole located at a distance greater than $r$ from the monopole center,
\begin{equation}
m_{\rm out}(r)=4\pi \int_r^\infty \rho(r') r'^2 dr' \,,
\label{eq18}
\end{equation}
must approach zero for sufficiently large values of $r$:
\begin{equation}
\alpha > 3/2 \Rightarrow \lim_{r \to \infty} m_{\rm out}(r) =0 \,.
\label{eq19}
\end{equation}
Therefore, $\alpha = 3/2$ defines a lower limit on $\alpha$ above which finite-energy global K-monopole solutions in Minkowski spacetime are possible.

An upper limit on the value of $\alpha$ may be obtained by requiring that $\rho \ge |p|$  at large distances from the global K-monopole core. This is a necessary condition for the dominant energy condition to hold (see \cite{Kontou:2020bta} for a review of energy conditions in general relativity and quantum field theory).  For $r \gg 1$ the proper energy density $\rho$ and the proper pressure $p$ are given approximately by $\rho=|X(r)|^\alpha $ and $p=(2 \alpha /3 - 1) |X(r)|^\alpha$, respectively, thus implying that 
\begin{equation}
    p (r) \sim \left(\frac23 \alpha - 1\right) \rho(r)\,.
\label{eq20}
\end{equation}
This then leads to the following upper limit on $\alpha$:
\begin{equation}
w \equiv \frac{p}{\rho} \sim \frac{2}{3} \alpha - 1 \leq 1 \Rightarrow \alpha \leq 3\,.
\label{eq21}    
\end{equation}
Thus, static global K-monopole solutions with finite energy satisfying the condition $\rho \ge |p|$ exist only for $\alpha \in (3/2, 3]$.

\section{Von Laue Condition \label{sec3}}

The existence of stable compact objects in Minkowski spacetime requires their volume-averaged internal pressure to be zero, a constraint known as the von Laue condition (see \cite{Laue:1911} for von Laue's original derivation, as well as \cite{Avelino:2018qgt,Giulini:2018tuw,Avelino:2023rac} for alternative approaches). Here, we present a derivation of this condition, following closely Giulini’s reasoning \cite{Giulini:2018tuw} applied to a covariantly conserved energy-momentum tensor. 

In Minkowski spacetime, for a time-independent energy-momentum tensor, one has
\begin{equation}
\partial_j T^{\mu j} = 0.
\label{eq22}
\end{equation}
Consider a volume $\Sigma$ within a spacelike affine hyperplane of Minkowski space, defined by a fixed physical time $t$, and its boundary $\partial \Sigma$. Using the identity
\begin{equation}
\partial_j (T^{\mu j}x^i) = (\partial_j T^{\mu j})x^i+ T^{\mu j}\partial_j x^i
\label{eq23}
\end{equation}
together with Eq. (\ref{eq22}) and Stokes's theorem, one obtains
\begin{equation}
\int_\Sigma T^{\mu j}\delta_j^i d^3x= \int_{\partial\Sigma} T^{\mu j} x^i n_j dS\,,
\label{eq24}
\end{equation}
where, $n_j$ are the covariant components of the outward-pointing normal to $\partial\Sigma$. If  $T^{\mu j} = 0$ at $\partial \Sigma$ or if $\Sigma$ represents the whole fixed time hyperplane and $T^{\mu j}$ decays sufficiently fast (e.g., if $T^{\mu j} \propto 1/r^{3+\epsilon}$ with $\epsilon > 0$) then
\begin{equation}
\int_\Sigma T^{\mu i} d^3x= 0\,.
\label{eq25}
\end{equation}
Equation (\ref{eq25}) implies that the spatial components of a particle’s 4-momentum, defined as
\begin{equation}
P^{i} \equiv\int_\Sigma T^{0 i} d^3x= 0\,,
\label{eq26}
\end{equation}
vanish in the chosen frame of reference (the particle's proper frame).  Moreover, Eq. (\ref{eq25}) is necessary and sufficient for 
$P^\mu$ to transform as a four-vector under Lorentz transformations \cite{Giulini:2018tuw}. The von Laue condition also follows directly from Eq. (\ref{eq25}),
\begin{equation}
\frac13\int_\Sigma  T^{ii} d^3x = \int_\Sigma  p d^3x= 0\,.
\label{eq27}
\end{equation}
For $\alpha > 3/2$ a global K-monopole possesses finite energy. In this case, 
\begin{equation}
\bar{w}(r) \equiv \frac{ \langle p\rangle_r}{\langle \rho \rangle_r}\,
\label{eq28}
\end{equation}
must approach zero sufficiently far from its center,
\begin{equation}
\lim_{r \to \infty} \bar w (r) = 0\,.
\label{eq29}
\end{equation}
Here,
\begin{equation}
\langle \mathcal V \rangle_r = \frac{3 \int_0^r \mathcal V(r') r'^2 dr'}{r^3}
\label{eq230}
\end{equation}
represents the volume average of the physical variable $\mathcal V$ inside a sphere of radius $r$ centered on the global K-monopole.

The condition $\mathcal{L}_{\rm on-shell}  = -\rho$ [see Eq. (\ref{eq12})], valid in the particle's proper frame, in combination with Eq. (\ref{eq27}), then implies that the volume average of the trace of the energy-momentum tensor, $T \equiv T^{\mu}_{\ \ \mu}$, of a static compact object of fixed proper mass and structure with a negligible self-induced gravitational field (e.g., a global K-monopole with $\alpha > 3/2$) is equal to the volume average of its on-shell Lagrangian,
\begin{equation}
\int_\Sigma T d^3x= \int_\Sigma (-\rho +3p)d^3x=\int_\Sigma \mathcal{L}_{\rm on-shell} d^3x\,.
\label{eq31}
\end{equation}
Thus, the ratio of the volume-averaged on-shell Lagrangian to the volume-averaged trace of the energy-momentum tensor must approach unity for sufficiently large values of $r$,
\begin{equation}
\lim_{r \to \infty} \frac{\langle \mathcal{L}_{\rm on-shell}  \rangle_r} {\langle T  \rangle_r}  = 1\,.
\label{eq32}
\end{equation}
Note that not only is $\mathcal{L}_{\rm on-shell}=-\rho$ (in the particle's proper frame) a sufficient condition for recovering the action of a free particle ($S=-m\int d\tau$, where $\tau$ is its proper time) after integration over the spatial volume, but it is also a necessary condition (at least in an average sense).

\section{Global K-monopole: numerical methods \label{sec4}}

In this section, we provide a brief description of the analytical results and numerical methods used to obtain static global K-monopole solutions.

To solve the differential equation given in Eq. (\ref{eq7}), we employed the shooting method (see, for example, \cite{Press2007}). This approach involves specifying boundary conditions at a point near the global K-monopole center ($r = r_0 \sim 0$) and integrating the field equations outward to a larger radius, $r = r_1$. For $r \ll 1$, the function $f(r)$ can be approximated as follows \cite{Babichev_2006}:
\begin{equation}
f(r) = Ar - \left( \frac{2}{3A^2} \right)^{\alpha - 2} \frac{\lambda r^3}{5A \alpha(1 + 2\alpha)}  + O(r^4)\,,
\label{eq33}
\end{equation}
where $A$ is an unknown constant. Using the shooting method, we iteratively adjusted $A$ by modifying the values of $f(r_0)$ and $f'(r_0)$ to ensure that $f(r_1)$ would approach unity. This process employed a binary search algorithm. Two initial estimates of $A$ were chosen: one that undershot the boundary condition at $r=r_1$ [$f(r_1)<1$] and another that overshot it [$f(r_1)>1$]. In each iteration, the midpoint of the interval defined by these two estimates of $A$ was computed. Depending on whether the resulting $f(r_1)$ was  above or below unity, the interval was halved, retaining only the subinterval containing the correct value of $A$. This process was repeated several times until the desired level of accuracy was achieved.

In the tail region of the global K-monopole ($r \gg 1$), the field equation [Eq. (\ref{eq7})] can be approximated as
\begin{equation}
\frac{2}{r^2}f + \frac{1}{\alpha|X|^{\alpha-1}}\frac{dV}{df} = 0\,,
\label{eq34}
\end{equation}
where $|X| \sim f^2/r^2$. Substituting this into Eq. (\ref{eq34}) yields
\begin{equation}
f^{2\alpha-2}+\frac{\lambda r^{2\alpha}}{2\alpha}(f^2-1)=0\,.
\label{eq35}
\end{equation}
The function
\begin{equation}
f(r) = 1 - \frac{\alpha}{\lambda} r^{-2\alpha}
\label{eq36}
\end{equation}
provides a solution to Eq. (\ref{eq35}) and, therefore, to the complete field equations at large distances from the monopole core. The full solution $f(r)$ of the second-order differential field equation [Eq. (\ref{eq7})] was obtained by combining the shooting method for the core and near-core regions with the asymptotic approximation [Eq. (\ref{eq36})] for $r \gg 1$. 

To validate our numerical results, we also used the relaxation method (see, e.g., \cite{Press2007}).  This technique solves the field equation [Eq. (\ref{eq7})] by iteratively refining an initial guess for $f(r)$ until convergence, while imposing the appropriate boundary conditions. The results were consistent with the ones obtained using the shooting method.

%%%%%%%%%%%%%%%%%%%%%%%
\begin{figure}[t!]
\centering
\includegraphics[width=\columnwidth]{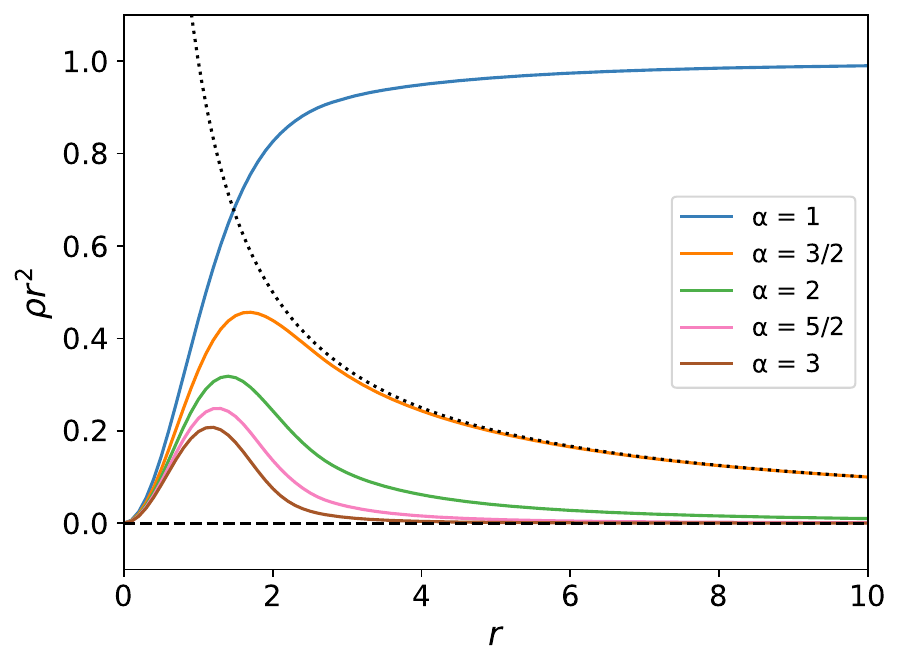}
\caption{The solid lines show the value of $\rho  r^2$ as a function of the distance to the monopole center ($r$) for static global K-monopole solutions obtained numerically considering various values of $\alpha$ (the dashed black line represents the vacuum solution with $\rho=0$ and the dotted line represents the function $1/r$). For $\alpha=1$, $\rho r^2$ tends to unity far from the monopole center, while for $\alpha > 1$ it tends to zero in that limit. Also notice that if $\alpha= 3/2$ (the threshold above which the global K-monopole solutions have finite energy) $\rho r^2 \sim  1/r$ for $r \gg 1$. 
\label{fig1}
}
\end{figure}

\section{Global K-monopole: numerical results \label{sec5}}

In this section, we present and discuss global K-monopole solutions. These solutions are obtained by numerically solving Eq. (\ref{eq7}) with the appropriate boundary conditions, using the methods outlined in Sec. \ref{sec4}. We shall characterize the structure of global K-monopoles through the determination of the dependence of key physical variables --- the proper energy density $\rho$, the proper pressure $p$, the equation of state parameters $w$ and $\bar w$, and $\langle \mathcal{L}_{\rm on-shell}  \rangle_r /\langle T  \rangle_r$ --- on the radial distance ($r$) to the monopole center for different values of $\alpha$. The analysis focuses on the parameter range $\alpha \in (3/2, 3]$, where both the von Laue and the condition $\rho \ge |p|$ are satisfied [Eqs. (\ref{eq19}) and (\ref{eq21}), respectively]. When relevant, we also consider the standard global K-monopole case ($\alpha = 1$) and the infinite-energy limit ($\alpha = 3/2$). For definiteness, we set $\lambda = 1$, noting that this choice does not affect our main results.

%%%%%%%%%%%%%%%%%%%%%%%
\begin{figure}[t!]
\centering
\includegraphics[width=\columnwidth]{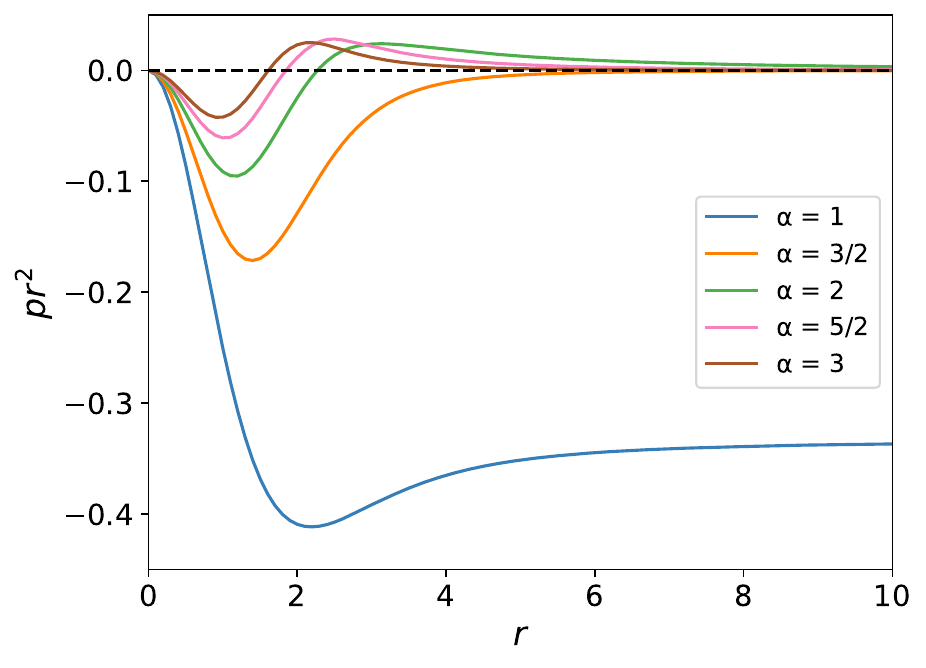}
\caption{The solid lines display the value of $p  r^2$ as a function of $r$ for various values of $\alpha$ (the dashed black line represents the vacuum solution with $p=0$). For $\alpha=1$, $pr^2$ tends to $-1/3$ for sufficiently large values of $r$, while for $\alpha>1$ it vanishes in that limit. Also notice that if $\alpha \leq 3/2$ the pressure is negative everywhere, while for $\alpha > 3/2$, the pressure is negative in the inner region and positive in the outer one.
\label{fig2}
}
\end{figure}

Figure \ref{fig1} shows the value of $\rho  r^2$ as a function of the distance to the monopole center $r$ for static global K-monopole solutions obtained numerically considering various values of $\alpha$ (solid lines). The dotted line represents the function $1/r$. This function provides an excellent approximation to $\rho r^2$ for $\alpha=3/2$ (the threshold above which the static global K-monopole solutions have finite energy) far from the monopole center ($r \gg 1$). Note that for $\alpha=1$, $\rho r^2$ tends to unity far from the global K-monopole center, while for $\alpha > 1$ it tends to zero in that limit (in the latter case it approaches the vacuum solution with $\rho=0$ defined by the dashed line). According to Eq. (\ref{eq16}), the proper energy density $\rho$ outside the core of the global K-monopole  decreases more rapidly for larger values of $\alpha$. This can be confirmed in Fig. \ref{fig1} which shows that the energy is more concentrated close to the monopole center for larger values of $\alpha$. Consequently, the monopole spatial interaction range should decrease with increasing $\alpha$.

Figure \ref{fig2} shows the value of $p r^2$ as a function of $r$ for static monopole solutions obtained numerically for different values of $\alpha$ (solid lines). For $\alpha=1$ the function $pr^2$ tends to $-1/3$ at large distances from the global K-monopole center, while for $\alpha>1$ it approaches zero in that limit (in the latter case it tends to the vacuum solution with $p=0$ defined by the dashed line). The pressure is negative close to the global K-monopole center for all values of $\alpha$. However, while for $\alpha \leq 3/2$ the pressure is negative for all values of $r$, for $\alpha > 3/2$, the pressure is negative in the inner region and positive in the outer one. Notice that the von Laue condition could not be satisfied if the pressure was negative everywhere.

The behavior of the pressure shown in Fig. \ref{fig2} resembles the pressure distribution inside a proton. The key difference lies in the sign of the pressure in the inner and outer regions. In the case of the proton, the pressure is positive close to its center (up to 0.6 fm) and negative at greater distances \cite{Polyakov:2018zvc, Shanahan:2018nnv, Burkert:2018bqq} (this is similar to the pressure profile of nuclei with baryon number $B=1$ in the Skyrme model \cite{GarciaMartin-Caro:2023toa}). In contrast, finite-energy global K-monopoles have a negative proper pressure in the inner region (this is also the case for all nuclei with $B>1$ in the Skyrme model \cite{GarciaMartin-Caro:2023toa}) and a positive proper pressure in the outer one. Despite the differences all of these systems satisfy the von Laue condition. 

%%%%%%%%%%%%%%%%%%%%%%%
\begin{figure}[t!]
\centering
\includegraphics[width=\columnwidth]{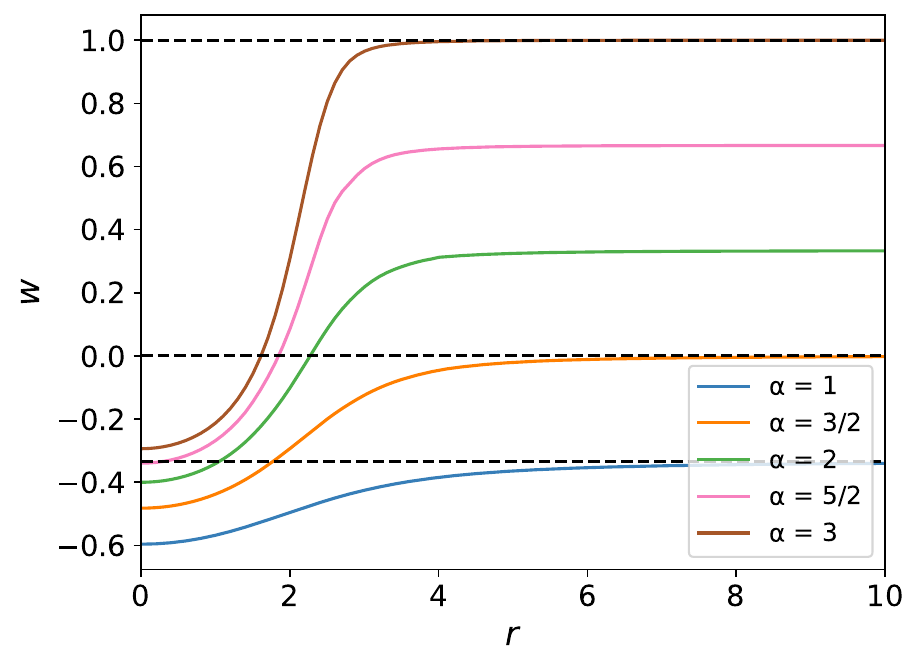}
\caption{The solid lines display the value of $w\equiv p/\rho$ as a function of $r$ for various values of $\alpha$, and the dashed black lines represent $w=1$, 0 and $-1/3$ (from top to bottom, respectively). At large distances from the monopole center, $w$ asymptotically approaches $2 \alpha/3 - 1$ [see Eq. (\ref{eq21})]. For $\alpha=1$, $w \to -1/3$; for $\alpha=3/2$, $w \to 0$; and for $\alpha=3$, $w \to 1$ (these limits are indicated by the dashed lines).}
\label{fig3}
\end{figure}

Figure \ref{fig3} shows the value of $w \equiv p/\rho$ as a function of $r$ for various values of $\alpha$ (solid lines). There are essentially two distinct regimes: one where $w$ increases with $r$, and another where it remains approximately constant --- the transition between these two regions is sharper for larger values of $\alpha$. Far from the monopole center, $w$ tends to $-1/3$ for $\alpha = 1$ (the standard case), to $0$ for $\alpha = 3/2$ (the threshold above which global K-monopole solutions have finite energy), and to 1 for $\alpha = 3$ (the threshold beyond which global K-monopole solutions no longer satisfy the condition $\rho \ge |p|$). The regime of interest, $\alpha \in (3/2, 3]$, corresponds to the range of $\alpha$ for which $w_{\infty} \equiv \lim_{r \to \infty} w(r \to \infty)=2 \alpha/3-1 \in (0, 1]$.

%%%%%%%%%%%%%%%%%%%%%%%
\begin{figure}[t!]
\centering
\includegraphics[width=\columnwidth]{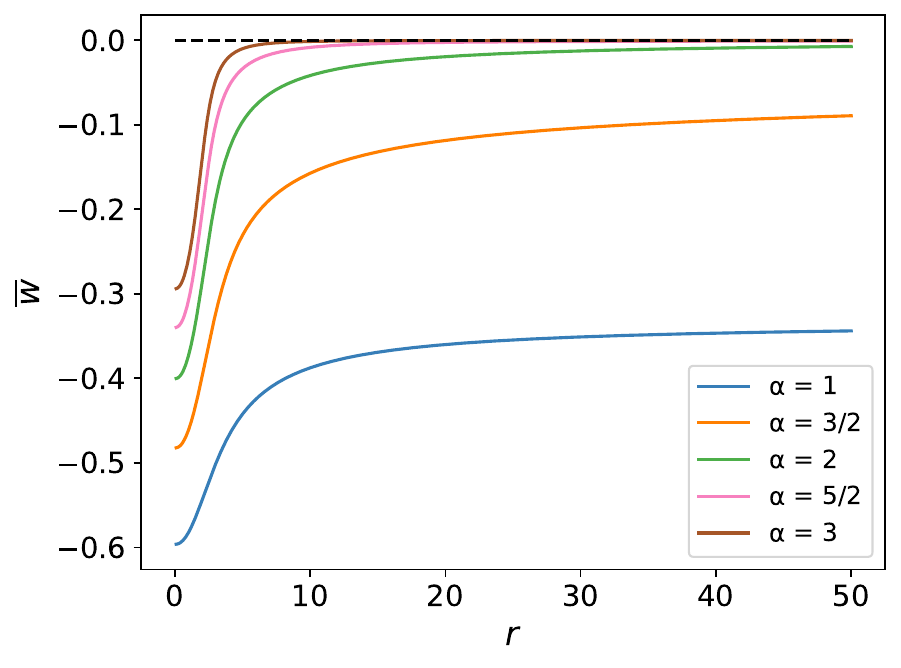}
\caption{The solid lines display the value of $\bar w\equiv \langle p \rangle_r/\langle \rho \rangle_r$ as a function of $r$ for various values of $\alpha$ (the dashed black line represents $\bar{w}=0$). Notice that for $\alpha > 3/2$ the value of $\bar{w}$ approaches zero at large distances from the global K-monopole center in agreement with the von Laue condition.
\label{fig4}
}
\end{figure}

Both the von Laue and the $\rho \ge |p|$ conditions are satisfied for $\alpha \in (3/2,3]$. These are equivalent, respectively, to the following two conditions: $\bar w_\infty\equiv \lim_{r \to \infty} \bar w (r) = 0$ and $|w| \le 1$ (everywhere). Hence, although $\bar w$ is generally nonzero (even for $r \gg 1$), such  deviations from zero should become negligible sufficiently far from the monopole center. Otherwise, the von Laue condition would not be satisfied. The dependence of $\bar w$ on $r$ is presented in Fig. \ref{fig4} for several values of $\alpha$ (solid lines). Figure \ref{fig4} confirms that, for $\alpha > 3/2$, $\bar{w}$ approaches zero (dashed line) for sufficiently large values of $r$,
\begin{equation}
\bar w (r \gg 1) \sim 0\,.
\label{eq37}
\end{equation}
This condition can be viewed as an approximate von Laue condition, which should apply to some degree even if the monopoles are not isolated (e.g. in the case of a network of global K-monopoles with $\alpha >3/2$, as long as the average between them is much larger than their core radius).

\begin{figure}[t!]
\centering
\includegraphics[width=\columnwidth]{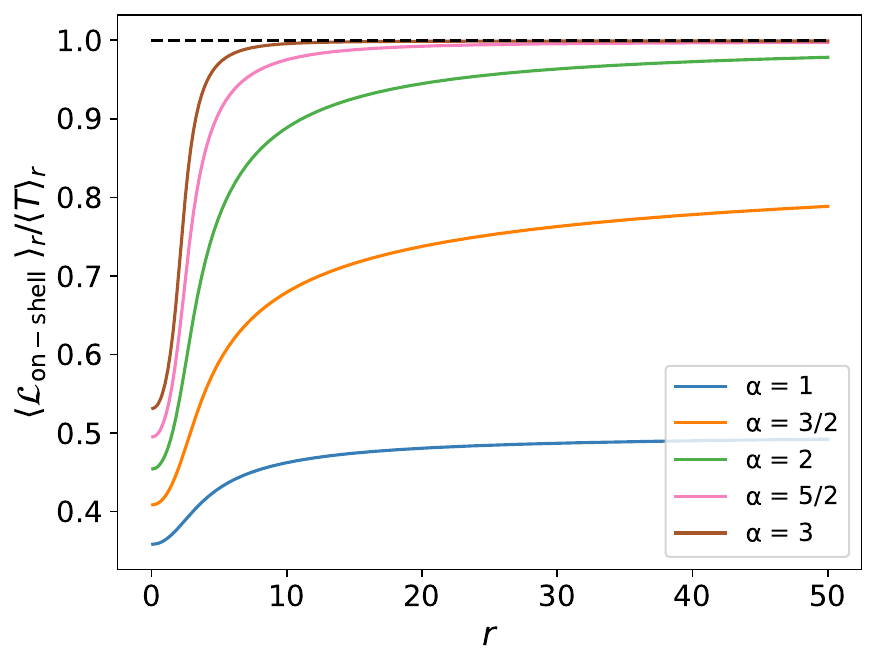}
\caption{The solid lines display the value of $\langle \mathcal{L}_{\rm on-shell}  \rangle_r/\langle T \rangle_r$ as a function of $r$ for various values of $\alpha$ (the dashed black line represents $\langle \mathcal{L}_{\rm on-shell}  \rangle_r/\langle T \rangle_r=1$). Notice that for values of $\alpha > 3/2$, at large distances from the monopole center, $\langle \mathcal{L}_{\rm on-shell}  \rangle_r$ approaches $\langle T \rangle_r$.
\label{fig5}
}
\end{figure}

Figure \ref{fig5} shows the value of $\langle \mathcal{L}_{\rm on-shell}  \rangle_r/\langle T \rangle_r$ as a function of $r$ for different values of $\alpha$ (solid lines). For values of $\alpha > 3/2$, far from the monopole center, $\langle \mathcal{L}_{\rm on-shell} \rangle_r/\langle T \rangle_r$ approaches unity (represented by the dashed line). This is expected since [see Eqs. (\ref{eq12}) and (\ref{eq14})]
\begin{equation}
\frac{\langle \mathcal{L}_{\rm on-shell}  \rangle_r}{\langle T \rangle_r} =\frac{-\langle\rho\rangle_r}{-\langle\rho\rangle_r+3\langle p \rangle_r} = \frac{1}{ 1-3\bar{w}(r)}\,,
\label{eq38}
\end{equation}
and $\bar{w}_\infty=0$.

\section{Real particles: deviations from the von Laue condition \label{sec6}}

In this section, we derive an upper limit for deviations from the von Laue condition, which arise because particles in realistic scenarios are not isolated. These deviations may be estimated by calculating the value of $\bar w$ at a distance $r$ from the particle center equal to the mean interparticle distance, defined by $d=n^{-1/3}$, where $n$ is the number density.

%%%%%%%%%%%%%%%%%%%%%%%

Employing the von Laue condition, the value of $\bar w$ can be written as a function of $r$ as
\begin{eqnarray}
\bar w(r) &\equiv&  \frac{\langle p \rangle_r}{\langle \rho \rangle_r} =\frac{\int_0^r p(r') r'^2 dr'}{\int_0^r \rho(r') r'^2 dr'}
= -\frac{\int_r^\infty p(r') r'^2 dr'}{\int_0^r \rho(r') r'^2 dr'} \nonumber \\
&=& - \frac{\int_r^\infty w(r') \rho(r') r'^2 dr'}{\int_0^r \rho(r') r'^2 dr'}\,.
\label{eq39}
\end{eqnarray}
For illustration purposes, we first consider the case of a global K-monopole.  As shown in Fig. \ref{fig3}, $w$ approaches $w_\infty$ sufficiently far from the core of an isolated global K-monopole. In this limit Eq. (\ref{eq39}) implies that:
\begin{eqnarray}
\bar w(r) &\sim&  -w_\infty \frac{\int_r^\infty \rho(r') r'^2 dr'}{\int_0^r \rho(r') r'^2 dr'}\nonumber \\
&=& -w_\infty \frac{m_{\rm out}(r)}{m-m_{\rm out}(r)}\,,
\label{eq40}
\end{eqnarray}
which simplifies to:
\begin{equation}
\bar w(r)\sim - w_\infty \frac{m_{\rm out}(r)}{m}\,,
\label{eq41}
\end{equation}
in the region where $w \sim w_\infty$ and $m_{\rm out}/m \ll 1$.

%%%%%%%%%%%%%%%%%%%%%%%
\begin{figure}[t!]
\centering
\includegraphics[width=\columnwidth]{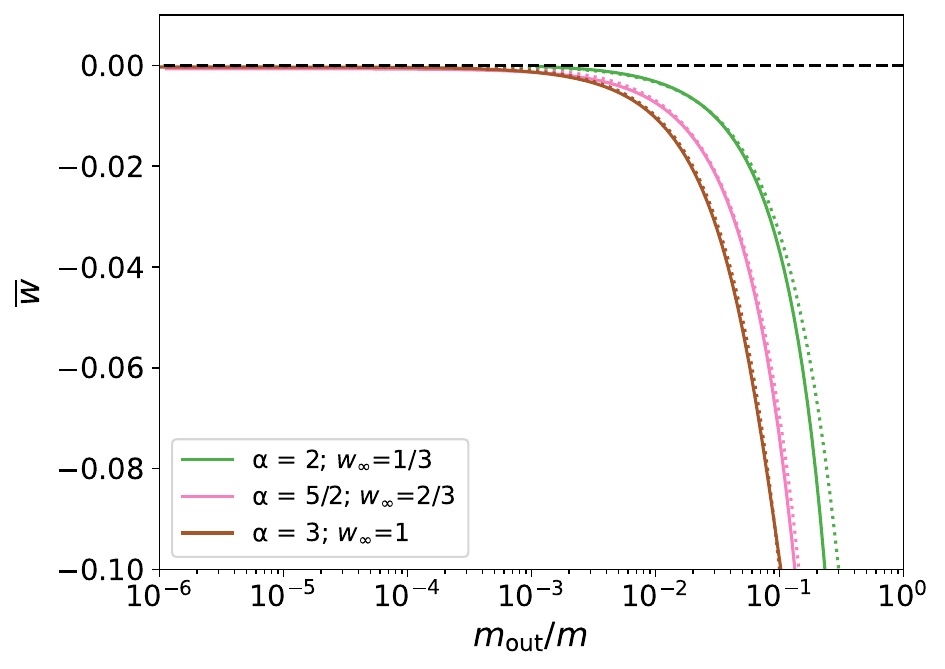}
\caption{The solid lines display the value of $\bar{w}$ as a function of $m_{\rm out}/m$ for various values of $\alpha$, and the dotted lines show the analytical approximation given in Eq. (\ref{eq41}) (the dashed black line represents $\bar{w}=0$). Notice that $\bar w$ becomes extremely close to zero for values of $m_{\rm out}/m \lsim 10^{-3}$.
\label{fig6}
}
\end{figure}

Figure \ref{fig6} shows the value of $\bar{w}$ as a function of $m_{\text{out}} / m$ (the fraction of the monopole mass  located outside a sphere of radius $r$ concentric with the monopole) for three different values of $\alpha$ (solid lines). The dotted lines depict the analytical approximation defined in Eq. (\ref{eq41}), which provides an excellent approximation to the numerical results provided that $m_{\rm out}/m \ll 1$. The differences between the curves obtained for $\alpha=2,\,5/2$, and $3$ are essentially due to the differences in $w_\infty=2\alpha /3 -1$.

In the case of real particles it is possible to obtain an upper limit on $\bar w (r)$ by considering the condition $\rho \ge |p|$  ($|w| \le 1$). Again, note that this condition is automatically satisfied if the dominant energy condition holds. Combined with Eq. (\ref{eq39}), this condition yields
\begin{equation}
|\bar w(r)| \le   \frac{\int_r^\infty \rho(r') r'^2 dr'}{\int_0^r \rho(r') r'^2 dr'} = \frac{m_{\rm out}(r)}{m-m_{\rm out}(r)} \,,
\label{eq42}
\end{equation}
which can be approximated as
\begin{equation}
|\bar w(r)| \le   \frac{m_{\rm out}(r)}{m}\,,
\label{eq43}
\end{equation}
if $m_{\rm out}/m \ll 1$.

%%%%%%%%%%%%%%%%%%%%%%%
\begin{figure}[t!]
\centering
\includegraphics[width=\columnwidth]{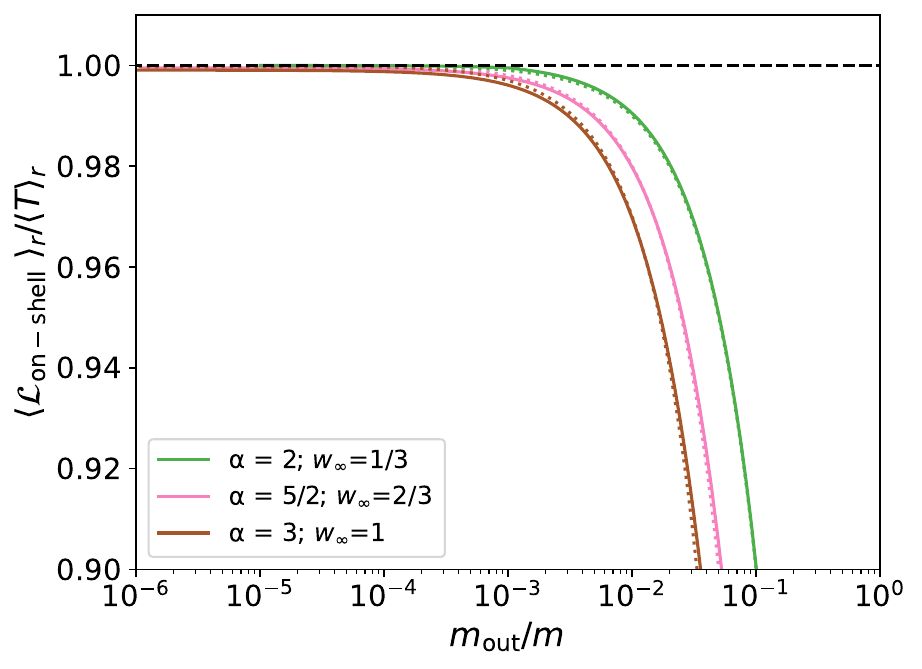}
\caption{The solid lines display the value of $\langle \mathcal{L}_{\rm on-shell}  \rangle_r/\langle T \rangle_r$ as a function of $m_{\rm out}/m$ for various values of $\alpha$, and the dotted lines show the analytical approximation given in Eq. (\ref{eq44}) (the dashed black line represents $\langle \mathcal{L}_{\rm on-shell}  \rangle_r=\langle T \rangle_r$). Notice that $\langle \mathcal{L}_{\rm on-shell}  \rangle_r/\langle T \rangle_r$ approaches unity for $m_{\rm out}/m \ll 1$.
\label{fig7}
}
\end{figure}

It is interesting to consider the implications of this result in the case of the proton. Although there are still significant experimental discrepancies between different measurements of the proton proper radius, we shall adopt the reference value of $r_{\rm proton} = 0.84 \, \mathrm{fm}$ \cite{Pohl:2010zza,Xiong:2019umf,Bezginov:2019mdi} (this choice does not qualitatively affect our results). Taking into account that the energy associated with the electrostatic field outside the proton ($r > r_{\rm proton}$) is approximately equal to $ 0.86 \, \mathrm{MeV}$, the ratio of the proton electric energy to its mass can be shown to be approximately equal to $10^{-3}$. Equation (\ref{eq43}) then implies that $|\bar w(r_{\rm proton})| \lsim 10^{-3}$ or, equivalently, that the von Laue condition applies within 0.1$\%$ inside the proton, here defined as the region with $r<r_{\rm proton}$ [in fact this upper limit on $|\bar w(r_{\rm proton})|$ can be further reduced by a factor of $1/3$, considering that $w=1/3$ for an electrostatic field]. This also holds for stable atomic nuclei with a baryon number $B$ greater than unity (in fact, the constraint will be stronger by a factor of approximately 2, given that neutrons carry no net electric charge).

Next, we examine the approximation $\langle \mathcal{L}_{\rm on-shell}  \rangle_r \sim \langle T \rangle_r$. For global K-monopoles with finite energy ($\alpha > 3/2$), Eqs. (\ref{eq38}) and (\ref{eq41}) yield
\begin{equation}
\frac{\langle \mathcal{L}_{\rm on-shell}  \rangle_r}{\langle T \rangle_r} \sim 1 - 3w_{\infty}\frac{m_{\text{out}}}{m}
\label{eq44}
\end{equation}
in the region where $w \sim w_\infty$ and $m_{\rm out}/m \ll 1$. This is confirmed in Fig. \ref{fig7},  where $\langle \mathcal{L}_{\rm on-shell}  \rangle_r/\langle T \rangle_r$ is plotted as a function $m_{\text{out}}(r) / m$ for three different values of $\alpha$ (solid lines) --- the dotted lines show the analytical approximation given in Eq. (\ref{eq44}) and the dashed black line represents the condition $\langle \mathcal{L}_{\rm on-shell}  \rangle_r/\langle T \rangle_r=1$. Figure \ref{fig7} shows that the analytical approximation provides an excellent approximation to the numerical results provided that $m_{\rm out}/m \ll 1$. The differences between the solid curves are associated to the dependence of $w_\infty$ on $\alpha$ ($w_\infty=2\alpha/3-1$).

Equation (\ref{eq38}) remains valid for real particles and the dominant energy condition is also expected to hold. Consequently, Eqs. (\ref{eq38}) and (\ref{eq42}) can be used to establish an upper limit on the deviation of $\langle \mathcal{L} \rangle_r/\langle T \rangle_r$ from unity,
\begin{equation}
\left|\frac{\langle \mathcal{L}_{\rm on-shell}  \rangle_r}{\langle T \rangle_r}-1\right| \lsim  3 \frac{m_{\rm out}(r)}{m}\,.
\label{eq45}
\end{equation}
For a proton, when $r> r_{\rm proton}$, this limit is even stricter (by a factor of 3), as $w=1/3$ for an electrostatic field. Considering that for a proton $m_{\rm out}(r_{\rm proton})/m \sim 10^{-3}$, the deviation of $\langle \mathcal{L}_{\rm on-shell}  \rangle_r / \langle T \rangle_r$ from unity at $r=r_{\rm proton}$ would not exceed $10^{-3}$. Hence, the condition $\langle \mathcal{L}_{\rm on-shell}  \rangle = \langle T \rangle$ applies within 0.1$\%$ inside the proton. Again, a somewhat stricter constraint applies to stable atomic nuclei with a baryon number $B$ greater than one.

\section{On-shell Lagrangian of real fluids \label{sec7}}

In this section, we examine how the results derived in Sec. \ref{sec6} determine the form of the on-shell Lagrangian of real fluids. We begin by focusing on minimally coupled fluids, after which we consider scenarios that involve nonminimal couplings to other matter fields or gravity.

The results in the previous section were established in the proper frame of individual particles. However, it is important to note that Eq. (\ref{eq45}) is also valid for moving particles. This is because both the Lagrangian and the trace of the energy-momentum tensor are scalar quantities, thus implying that their values at any given spacetime point remain invariant under Lorentz transformations. Moreover, the volume integrals of these quantities are evaluated over the same spacetime region.

Now, let us turn our attention to a real fluid made up of numerous moving particles. If the average distance between particles is much larger than their physical size (here defined as the proper radius that encloses most of the particle's mass), then the microscopic on-shell Lagrangian of the particle ensemble can, to an excellent approximation, be treated as the sum of the individual particle on-shell Lagrangians. This principle also holds for the trace of the energy-momentum tensor. Consequently, as long as spacetime can be approximated as Minkowskian at the smallest proper length scale where the fluid approximation is valid, the fluid’s Lagrangian (averaged over this scale) should effectively match the trace of its energy-momentum tensor. This relationship holds for a baryonic fluid as long as its proper energy density is smaller than that of atomic nuclei.

In general relativity, the energy and momentum of minimally coupled fluids are conserved. Once an equation of state $p(\rho)$ is specified,   along with appropriate boundary conditions, energy-momentum conservation completely determines the dynamics of a perfect fluid in Minkowski spacetime. This holds true regardless of the specific form of its on-shell Lagrangian. However, this independence breaks down when the perfect fluid interacts nonminimally with other fluids or fields.
A simple example of such a coupling involves modifying the minimally coupled particle Lagrangian $\mathcal L$ by multiplying it with an arbitrary regular function $\mathfrak f$ of a scalar field $\chi$. The Lagrangian of the corresponding fluid $\mathcal{L}_\text{F}$ is modified in the same manner. This results in the following modified Lagrangian:
\begin{equation}    \mathcal{L}_{\text{F}\chi}=\mathfrak f(\chi)\mathcal{L}_\text{F}\,.
\end{equation}
Thus, if $\mathcal{L}_\text{F}$ represents a fluid with particles of fixed mass $m$, then the modified Lagrangian $\mathcal{L}_{\text{F}\chi} =\mathfrak f(\chi)\mathcal{L}_\text{F}$ represents a fluid of particles of variable mass $M(\chi)$ such that $M(\chi) = \mathfrak f(\chi)m$.
In this scenario, the corresponding energy-momentum tensors $T_{\text{F}}^{\alpha\beta}$ and $T_{\text{F}\chi}^{\alpha\beta}$ are no longer conserved. Specifically, the conservation equation becomes \cite{Ferreira:2020fma}
\begin{equation}
    \nabla_\beta T_{\text{F}\chi}^{\alpha\beta} = \frac{\partial \mathfrak f}{\partial\chi}\mathcal{L}_\text{F}\partial^\alpha\chi \,.
\end{equation}
However, in many relevant scenarios (such as when $\chi$ represents a dark energy scalar field) the scalar field $\chi$ is nearly homogeneous on the cosmological frame and varies essentially on cosmological timescales. While it is important to note that the inhomogeneities of the scalar field can give rise to fifth forces between particles (see, for example, \cite{Ayaita:2011ay}), the effect of such a coupling on the profiles of the particle's on-shell Lagrangian $\mathcal{L}_{\rm on-shell}$ and of the energy-momentum trace $T$ is expected to be negligible.  Under such conditions, the results derived in the previous section should remain valid. Therefore, the condition $\mathcal{L}_{\rm F\chi} = T_{\rm F\chi}$ will essentially hold, provided that the proper energy density of the fluid is significantly smaller than the proper energy density within the particles themselves. 

\section{Conclusions}
\label{concl}

The von Laue condition is a fundamental constraint necessary for the stability of isolated, finite-mass systems in the Minkowski limit of general relativity. However, it does not generally hold when interactions extend beyond a particle’s characteristic size. In this paper, we have shown, using global K-monopoles as an illustrative model, that deviations from the von Laue condition due to interactions between particles are minimal, except in extremely dense environments.
	
To this end we began by considering the case of isolated particles, where we demonstrated that the maximum deviation of the volume integral of proper pressure from zero is tightly bounded by the energy fraction outside this volume, provided the dominant energy condition is satisfied. We also established that the maximum deviation from unity of the ratio between the volume averages of the on-shell Lagrangian and of the trace of the energy-momentum tensor cannot exceed three times the outer energy fraction. These results imply that, for stable atomic nuclei in environments with energy densities much smaller than those found in the interiors of neutron stars, this ratio remains extremely close to unity. 
	
Using these findings, we have shown that, except in extremely dense environments with an average energy density comparable to that of the particles themselves, the ideal gas on-shell Lagrangian provides a highly accurate approximation of the true on-shell Lagrangian, even for nonideal gases with non-negligible interparticle interactions. This is particularly relevant, since the on-shell Lagrangian of a significant fraction of the energy content of the Universe, including dark matter, baryons and photons, may be estimated using this approximation. This is especially important in astrophysical and cosmological scenarios involving nonminimal couplings to other matter fields or gravity, where the equations of motion often depend explicitly on the fluid’s on-shell Lagrangian.
	
We expect this study to offer new insights into the interplay between the microscopic behavior of particles and the macroscopic dynamics of fluids, deepening our understanding of these systems across a wide range of physical contexts.

\acknowledgments
We thank Francisco Lobo, Lara Sousa, Miguel Pinto, Rui Azevedo, Tiago Gonçalves, Vasco Ferreira, and our other colleagues of the Cosmology group at Instituto de Astrofísica e Ciências do Espaço for enlightening discussions. We acknowledge the support by Fundação para a Ciência e a Tecnologia (FCT) through the research Grants No. UIDB/04434/2020 and No. UIDP/04434/2020. This work was also supported by FCT through the R$\&$D project 2022.03495.PTDC - {\it Uncovering the nature of cosmic strings}.	
\bibliography{article}
\end{document}